\documentclass[11pt]{article}
\usepackage{times}
\usepackage{sidecap}
\usepackage{geometry}
\usepackage{floatrow}
\usepackage{float}
\usepackage{hyperref}
\usepackage{graphics,graphicx}
\usepackage{subcaption}
\usepackage{epsf,wrapfig}
\usepackage{booktabs}    
\usepackage{siunitx}     
\usepackage{caption}     
\usepackage{makecell}
\usepackage{tabularx}
\usepackage{array}
\usepackage{multirow}
\usepackage{amssymb}

\usepackage{natbib,bm}

\usepackage{paralist}
\let\olditem\item
\renewenvironment{thebibliography}[1]{%
  \section*{\refname}
  \let\par\relax
  \renewcommand{\item}[1][]{\olditem[\textbullet]}%
  \inparaenum}{\endinparaenum}

\newcounter{aff}

\DeclareGraphicsExtensions{.jpg,.pdf,.png,.eps}

\usepackage{xcolor}

\usepackage{tcolorbox}

\bibpunct{(}{)}{;}{a}{}{,}

\usepackage{soul}

\definecolor{darkcyan}{rgb}{0.0, 0.0, 0.7}

\definecolor{boxcolor}{RGB}{193, 229, 255}
\definecolor{figboxcolor}{RGB}{193, 229, 255}
\definecolor{boxcolorgray}{RGB}{227, 227, 227}
\definecolor{darkcyan}{rgb}{0.0, 0.0, 0.7}

\sethlcolor{boxcolorgray}

\def\gtrsim{\lower 2pt \hbox{$\, \buildrel {\scriptstyle >}\over
{\scriptstyle \sim}\,$}}
\def\lesssim{\lower 2pt \hbox{$\, \buildrel {\scriptstyle <}\over
{\scriptstyle \sim}\,$}}

\def\approxlt{\lower.2em\hbox{$\buildrel < \over \sim$}}
\def\approxgt{\lower.2em\hbox{$\buildrel > \over \sim$}}

\def\arcsec{$^{\prime\prime}$}

\def\be{\begin{equation}}
\def\ee{\end{equation}}
\def\bea{\begin{eqnarray*}}
\def\eea{\end{eqnarray*}}

\def\arcsec{^{\prime\prime}}

\def\jref@jnl#1{{\rm#1}}
\def\aj{\jref@jnl{AJ}}			
\def\araa{\jref@jnl{ARA\&A}}		
\def\apj{\jref@jnl{ApJ}}			
\def\apjl{\jref@jnl{ApJ}}		
\def\apjs{\jref@jnl{ApJS}}		
\def\ao{\jref@jnl{Appl.~Opt.}}		
\def\apss{\jref@jnl{Ap\&SS}}		
\def\aap{\jref@jnl{A\&A}}		
\def\aapr{\jref@jnl{A\&A~Rev.}}		
\def\aaps{\jref@jnl{A\&AS}}		
\def\azh{\jref@jnl{AZh}}			
\def\baas{\jref@jnl{BAAS}}		
\def\jrasc{\jref@jnl{JRASC}}	
\def\grl{\jref@jnl{Geophys.~Res.~Let.}}
\def\jgra{\jref@jnl{J.~Geophys.~Res.~(Atmos.)}}
\def\jgrp{\jref@jnl{J.~Geophys.~Res.~(Plan.)}}	
\def\memras{\jref@jnl{MmRAS}}		
\def\mnras{\jref@jnl{MNRAS}}		
\def\pra{\jref@jnl{Phys.~Rev.~A}}	
\def\prb{\jref@jnl{Phys.~Rev.~B}}	
\def\prc{\jref@jnl{Phys.~Rev.~C}}	
\def\prd{\jref@jnl{Phys.~Rev.~D}}	
\def\pre{\jref@jnl{Phys.~Rev.~E}}	
\def\prl{\jref@jnl{Phys.~Rev.~Lett.}}	
\def\psj{\jref@jnl{Planet.~Sci.~J.}}	
\def\pasp{\jref@jnl{PASP}}		
\def\pasj{\jref@jnl{PASJ}}		
\def\pnas{\jref@jnl{PNAS}}		
\def\qjras{\jref@jnl{QJRAS}}		
\def\skytel{\jref@jnl{S\&T}}		
\def\solphys{\jref@jnl{Sol.~Phys.}}	
\def\sovast{\jref@jnl{Soviet~Ast.}}	
\def\ssr{\jref@jnl{Space~Sci.~Rev.}}	
\def\zap{\jref@jnl{ZAp}}			
\def\nat{\jref@jnl{Nature}}		
\def\iaucirc{\jref@jnl{IAU~Circ.}}
\def\aplett{\jref@jnl{Astrophys.~Lett.}}
\def\apspr{\jref@jnl{Astrophys.~Space~Phys.~Res.}}
\def\bain{\jref@jnl{Bull.~Astron.~Inst.~Netherlands}}
\def\fcp{\jref@jnl{Fund.~Cosmic~Phys.}}
\def\gca{\jref@jnl{Geochim.~Cosmochim.~Acta}}
\def\grl{\jref@jnl{Geophys.~Res.~Lett.}}
\def\jcp{\jref@jnl{J.~Chem.~Phys.}}	
\def\jgr{\jref@jnl{J.~Geophys.~Res.}}	
\def\jqsrt{\jref@jnl{J.~Quant.~Spec.~Radiat.~Transf.}}
\def\memsai{\jref@jnl{Mem.~Soc.~Astron.~Italiana}}
\def\nphysa{\jref@jnl{Nucl.~Phys.~A}}
\def\physrep{\jref@jnl{Phys.~Rep.}}
\def\physscr{\jref@jnl{Phys.~Scr}}
\def\planss{\jref@jnl{Planet.~Space~Sci.}}
\def\procspie{\jref@jnl{Proc.~SPIE}}

\def\arcsec{\hbox{$^{\prime\prime}$}}

\def\ion#1#2{#1$\;${\small\rm\@Roman{#2}}\relax}

\newcount\lecurrentfam
\def\LaTeX{\lecurrentfam=\the\fam \leavevmode L\raise.42ex
\hbox{$\fam\lecurrentfam\scriptstyle\kern-.3em A$}\kern-.15em\TeX}
\def\sizrpt{
(\fontname\the\font): em=\the\fontdimen6\font, ex=\the\fontdimen5\font
\typeout{
(\fontname\the\font): em=\the\fontdimen6\font, ex=\the\fontdimen5\font
}}
\definecolor{dark}{rgb}{ 0.34, 0.09, 0.49}

\usepackage{fancyhdr,url}
\usepackage{transparent}
\usepackage{tablefootnote}
\usepackage{mathpazo}
\begin{document}

\noindent
\begin{raggedright}
{\Huge\bfseries ALMA2040: Submillimeter-wave Limb Sounding of Venus and Mars \\[0.5em]
}\end{raggedright}

\noindent

Hideo Sagawa$^{1, 2}$, Arianna Piccialli$^3$, Miriam Rengel$^4$, Shohei Aoki$^{5, 6}$, Thibault Cavali\'e$^7$, Alessandra Migliorini$^8$, Raphael Moreno$^9$, Arielle Moullet$^{10}$

\vspace{1em}
{
\footnotesize{
\noindent $^1$ Kyoto Sangyo University, Kyoto 603-8555, Japan, \\
$^2$ Koyama Astronomical Observatory, KSU, Kyoto 603-8555, Japan, \\
$^3$ Belgian Institute for Space Aeronomy, BIRA-IASB, Brussels, Belgium, \\
$^4$ Max-Planck-Institut f\"ur Sonnensystemforschung, 37077 G\"ottingen, Germany, \\
$^5$ Graduate School of Frontier Sciences, The University of Tokyo, Kashiwa 277-8561, Japan, \\
$^6$ Graduate School of Science, Tohoku University, Sendai 980-0845, Japan, \\
$^7$ Laboratoire d'Astrophysique de Bordeaux, CNRS, UMR 5804, 33600 Pessac, France, \\
$^8$ Astronomical Observatory of Padova, OAPd-INAF, Italy,\\
$^9$ LIRA, Observatoire de Paris, 92195 Meudon Cedex, France,\\
$^{10}$ NRAO, Charlottesville, VA 22903-2475, USA.\\}
}
\begin{tcolorbox}[boxsep=3pt,left=0pt,right=0pt,top=1pt,bottom=1pt,colframe=boxcolor,colback=boxcolor]
\justify
\textbf{Abstract:} 
Venus and Mars are unique laboratories for understanding the evolution and diversity of terrestrial planets. Despite their broadly similar rocky compositions, their atmospheres, climates, and surface environments have evolved into dramatically different states. Understanding the physical and chemical processes responsible for these divergent pathways is essential for interpreting terrestrial exoplanets and their habitability. \\
By the 2040s, spacecraft missions will provide unprecedented close-up observations of Venus and Mars, while ground-based facilities can provide the long-term monitoring needed to place these snapshots in the context of atmospheric variability and planetary-scale circulation. ALMA2040 can uniquely bridge these two approaches by combining high-resolution submillimeter heterodyne spectroscopy with routinely accessible, extremely high angular resolution. \\
In particular, high-resolution limb sounding would enable vertically resolved measurements of atmospheric winds, temperature, and trace species from the ground. An angular resolution of approximately 0.01\arcsec~would allow spatial scales comparable to the atmospheric scale heights of Venus ($\sim$5~km) and Mars ($\sim$11~km) to be resolved. This capability would enable three-dimensional studies of Venusian upper-atmospheric circulation and Martian global dust storms, as well as systematic investigations of CO$_2$ photochemistry, water loss through HDO/H$_2$O fractionation, and Venusian sulfur chemistry and cloud formation. \\
By repeatedly observing different local times, seasons, and solar-activity conditions, ALMA2040 would extend spacecraft measurements from snapshots to long-term four-dimensional characterization. It would therefore provide a unique observational framework for directly connecting spacecraft discoveries with the temporal and spatial evolution of planetary atmospheres. 
\end{tcolorbox}

\newpage{}

\vspace{-0.7cm}
\section{Venus and Mars Observations in the 2040s}
\subsection{Scientific Rationale for Studying the Atmospheres of Venus and Mars}
Studies of Solar System bodies provide a unique pathway to understanding the formation and evolution of the Solar System and the conditions that made Earth habitable. By characterizing planetary environments and the physical and chemical processes that shape them, planetary science addresses fundamental questions extending beyond astronomy into geoscience, atmospheric science, and biology. A key strength of Solar System studies is the ability to conduct close-up and in-situ observations using spacecraft, including landers and sample-return missions. Such measurements provide high-resolution, accurate data that are essential for refining theoretical models and numerical simulations. 

Among the diverse research topics in planetary science, Venus and Mars are particularly valuable comparative laboratories for understanding the evolution of terrestrial planets. Despite sharing a rocky composition with Earth, they exhibit dramatically different surface environments in terms of atmospheric composition and mass, water abundance, and temperature. Understanding the origins of these differences is essential for identifying the physical and chemical processes that govern atmospheric evolution, water distribution, and climate on terrestrial planets.

This challenge is increasingly relevant in the era of exoplanet exploration. As the number of known exoplanets grows and proposed facilities such as the Habitable Worlds Observatory (HWO) enable more detailed studies of terrestrial worlds, Venus and Mars will provide essential benchmarks for interpreting distant planetary systems. Detailed atmospheric characterization of these neighboring planets can help distinguish processes common to terrestrial planets from those shaped by individual planetary properties and evolutionary histories. Venus and Mars are thus not only compelling targets in their own right but also key to understanding the diversity and habitability of terrestrial planets across the universe.

\subsection{Role of Ground-Based Observations in the 2040s}
By the 2040s, new generations of spacecraft are expected to have substantially advanced our understanding of Venus and Mars. At Venus, EnVision \citep{EnVision2023}, VERITAS \citep{Smrekar2022}, and DAVINCI \citep{Garvin2022} are expected to provide new insights into geological history, water evolution, atmospheric composition and chemistry, and present-day geological and atmospheric processes \citep{Limaye2023}. Mars exploration will likewise continue through orbiters, landers, and rovers, alongside growing interest in human exploration and resource utilization. These missions will provide unprecedented detail on the surface and atmospheric environments of both planets.

In this context, ground-based observations will play a role beyond complementing spacecraft measurements. Their key advantage is the ability to conduct repeated, long-term monitoring, providing the temporal baseline needed to characterize atmospheric variability beyond the duration of individual missions. Planetary atmospheres vary on timescales ranging from sub-hourly and diurnal cycles to seasonal and decadal changes. Long-term observations are therefore essential for identifying processes that emerge only over extended periods, as illustrated by recent evidence for decadal-scale climate change in the atmosphere of Venus \citep{Lee2019, Ando2025}.

Another major challenge is bridging the spatial scales sampled by spacecraft and ground-based observations. Atmospheric phenomena span localized structures to zonal, hemispheric, and global circulation patterns, while varying with altitude as different physical and chemical processes dominate. Photochemistry, atmospheric waves, and radiative processes, in particular, can strongly influence mass and energy transport in the upper atmosphere. Because ground-based observations typically have coarser spatial resolution than spacecraft measurements, they may probe different physical regimes, complicating direct comparisons between the two datasets.

A major goal for planetary observations in the 2040s should therefore be to establish a coordinated observing system that links spacecraft measurements with ground-based monitoring in both space and time. Ground-based facilities with spatial resolutions approaching those of spacecraft could track localized phenomena identified during spacecraft encounters over much longer periods and place them in the context of planetary-scale circulation and transport. This capability would bridge the gap between detailed spacecraft snapshots and the long-term evolution of planetary atmospheres.

\section{Achieving Spacecraft-Comparable Observations from the Ground}
ALMA2040 is expected to provide a new observational capability for planetary atmospheres by combining unprecedented angular resolution with the ability to conduct long-term, repeated observations from the ground. This combination will enable a direct connection between detailed spacecraft observations---snapshots of an atmosphere at a particular time and location---and sustained ground-based monitoring over a wide range of temporal and spatial scales.

Heterodyne spectroscopy at submillimeter wavelengths provides a powerful means of probing planetary atmospheres: for detail, see the review papers by \citet{Lellouch2008} and \citet{Cordiner2024}. Its high spectral resolution enables detailed measurements of molecular line profiles, allowing not only the abundance of trace species but also the physical state of the atmosphere, including temperature and winds, to be constrained. For Venus and Mars, molecular species such as CO, H$_2$O, HDO, SO, and SO$_2$ can be used to probe atmospheric circulation, transport, photochemistry, and the planetary water cycle. The scientific capabilities of submillimeter interferometry have already been demonstrated with the current ALMA, including measurements of the spatial distributions and short-term variability of SO, SO$_2$, and HDO \citep{Encrenaz2015,Piccialli2017} and the search and following verification tests for PH$_3$ \citep{Greaves2021,Villanueva2021} in the Venusian atmosphere, as well as atmospheric temperature and wind fields on Mars using CO \citep{Parks2021,Kakinuma2025}.

A particularly transformative capability of ALMA2040 will be to combine this high spectral resolution with sufficiently high angular resolution to enable high-resolution limb sounding of planetary atmospheres.
In limb sounding, the line of sight passes tangentially through the atmosphere rather than viewing the planetary disk from above, as in nadir observations. The resulting long atmospheric path length greatly enhances the sensitivity to trace species. Moreover, the strongest contribution to the observed signal comes from the tangent altitude, where the line of sight passes through the atmosphere at its closest approach to the planet. By spatially scanning the limb with sufficiently high angular resolution, observations at different tangent altitudes can therefore be obtained, providing direct information on the vertical structure of the atmosphere.

\begin{figure}[t]
\centering
\includegraphics[width=0.9\textwidth]{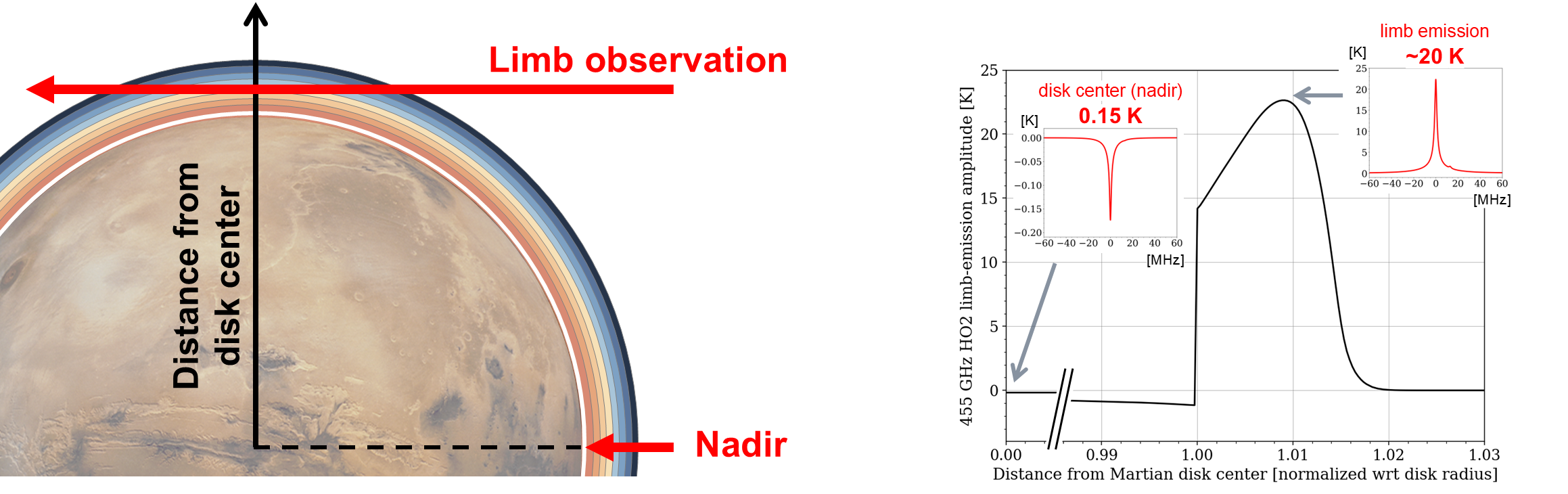}
\caption{\textbf{Schematic illustration of atmospheric limb observations.} The lines-of-sight for nadir and limb observations are shown on the left. The amplitudes of the spectral features, either in absorption or emission, expected for Martian HO$_2$ at 455.794~\si{\giga\hertz} are shown on the right. An HO$_2$ abundance of 0.2~ppbv is assumed in the calculation. No beam convolution is applied.}\label{fig:limbgeometry}
\end{figure}

Figure~\ref{fig:limbgeometry} shows an example of limb observations of a planetary atmosphere. The brightness-temperature difference from the continuum level is shown. In this example, we simulated observations of the HO$_2$ 455.794~\si{\giga\hertz} line at Mars. With the nadir-viewing geometry, the HO$_2$ spectral feature appears in absorption against the bright continuum emission from the planetary disk, with an absorption depth of only 0.15~K in brightness temperature. In contrast, in the limb-viewing geometry, the spectral feature appears in emission against the cold sky, and its amplitude increases significantly to $\sim$20~K because of the long line-of-sight path through the atmosphere. 
These results are without beam convolution. 
Compared to Mars' radius of 3396~km, the thickness of its atmosphere is at most 100~km, that is, less than 3$\%$. Although the apparent radius varies from a few arcseconds to a maximum of about 12\arcsec, the thickness of the atmosphere is still just under 3$\%$ of that. As the beam size increases, such a thin limb-emission signal is diluted and can become buried beneath the bright continuum emission from the planetary disk. 

The scientific utility of submillimeter limb sounding has already been demonstrated by space-based observations. Measurements by instruments such as the Microwave Limb Sounder (MLS), Odin Sub-Millimetre Radiometer (SMR), and the Superconducting Submillimeter-Wave Limb Emission Sounder (SMILES) have made major contributions to studies of terrestrial atmospheric chemistry and dynamics \citep[e.g.,][]{Baron2013,Kreyling2013}. The feasibility of limb observations from the ground has also been demonstrated with the current ALMA. In particular, trace-species emission has been detected in ring-like structures around the disks of Titan \citep{Cordiner2019}, whose atmosphere is extended relative to the satellite itself, and Jupiter \citep{Cavalie2023}, whose large planetary radius provides a sufficiently extended atmospheric limb. These observations demonstrate that ALMA can already probe planetary atmospheres along tangential lines of sight in geometrically favorable cases where the atmosphere is sufficiently thick. 

The key scientific goal for ALMA2040 is to extend this capability from a limited set of favorable targets and observing configurations to high-resolution, routine, and broadly applicable limb sounding of Venus and Mars.
With sufficiently high angular resolution, limb observations can spatially resolve different tangent altitudes and thereby transform the atmosphere from a two-dimensional projected disk into a vertically resolved structure. In particular, achieving spatial scales comparable to those sampled by spacecraft limb sounders would allow localized atmospheric structures and chemical features identified by spacecraft to be followed from the ground over extended periods.

Spacecraft limb observations can obtain broad spatial coverage as the spacecraft orbits a planet, but measurements at any given location are necessarily limited in time and often provide snapshots of the atmospheric state. ALMA2040, in contrast, could map the entire planetary limb in a single integration while repeatedly observing the same altitude ranges under different local times, seasons, and solar-activity conditions. This capability would make it possible to determine how localized phenomena discovered by spacecraft evolve within planetary-scale atmospheric circulation and seasonal variability, and to establish their long-term temporal evolution.

\begin{wrapfigure}{r}{70mm}
\vspace{-0.8cm}
 \begin{center}
 \includegraphics[width=0.99\linewidth]{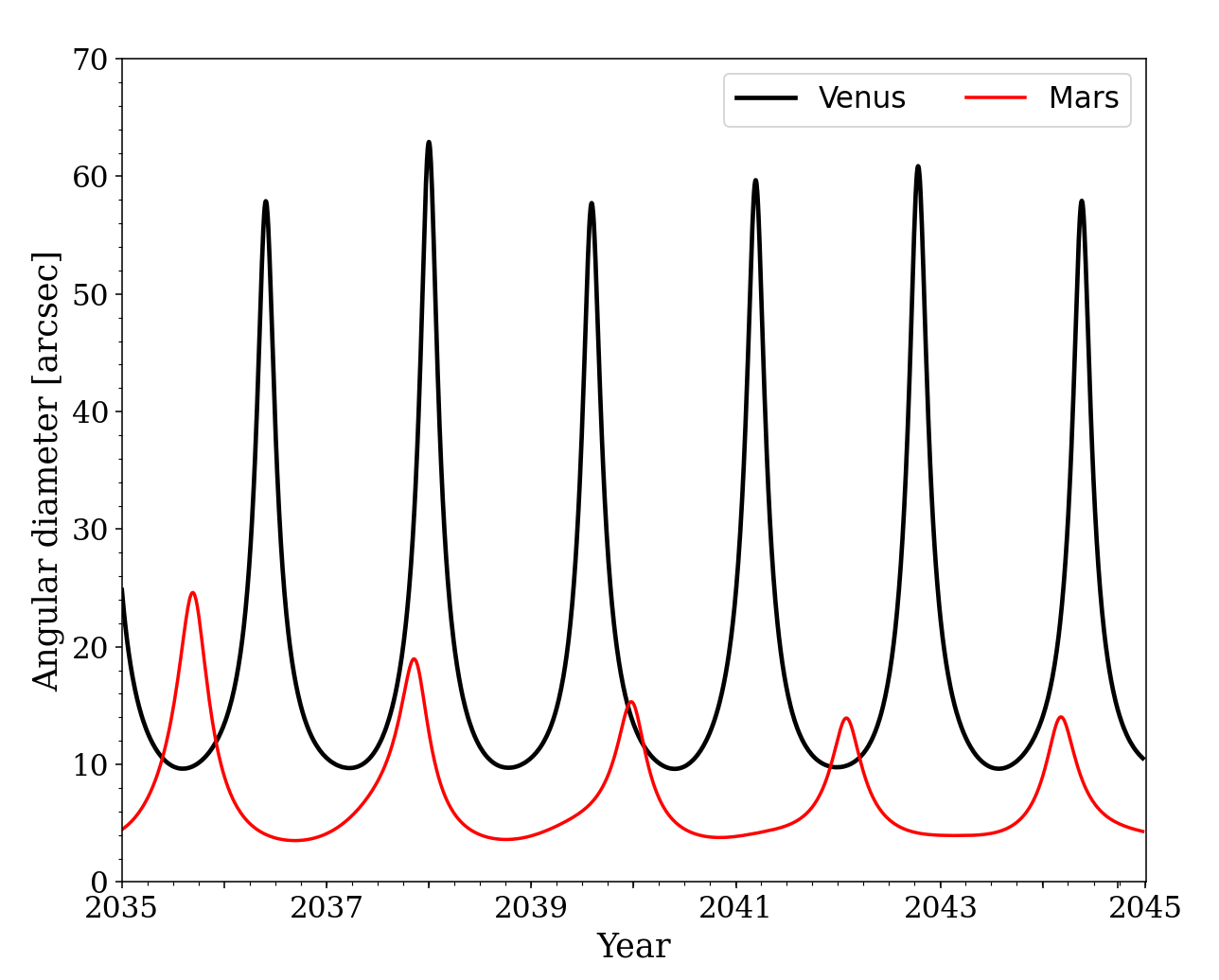}
 \caption{\textbf{Apparent angular sizes of Venus and Mars as a function of time.}}
 \label{fig:disksize}
 \end{center}
\end{wrapfigure}

A key requirement for such observations is spatial resolution comparable to the atmospheric scale height, which indicates the characteristic vertical structure of a planetary atmosphere. Representative values are approximately 5~km in the middle atmosphere of Venus and 11~km in the atmosphere of Mars. The apparent diameter of Venus varies from approximately 10\arcsec~to 60\arcsec~as its distance from Earth changes (Fig.~\ref{fig:disksize}). 
By setting the goal to achieve a 5~km limb-resolution during the period when both day and night local times on Venus can be observed at the eastern and western limbs (apparent diameter of approximately 25\arcsec), this corresponds to a spatial resolution of 0.01\arcsec. Mars has an apparent diameter ranging from a few arcseconds to approximately 15$-$25\arcsec~near closest approach; an angular resolution of approximately 0.01\arcsec~would likewise enable spatial scales comparable to its atmospheric scale height to be resolved.

Such an angular resolution of 0.01\arcsec~is already achievable with the current ALMA under limited observing conditions, such as the longest-baseline C43-10 configuration in Bands 8$-$10. The key requirement for ALMA2040, however, is not simply to improve the ultimate angular-resolution limit. For long-term monitoring of Venus and Mars, such spatial resolution must become routinely accessible rather than being restricted to specific array configurations or observing periods. Moreover, because scientifically important molecular tracers are distributed across different frequency bands, high angular resolution should be available not only in Bands 8$-$10 but also in lower-frequency bands, including Bands 5 and 6.
Thus, the scientific requirement for ALMA2040 is not merely higher maximum angular resolution. Rather, it is the capability to perform high-sensitivity, high-spectral-resolution submillimeter spectroscopy at scale-height-level spatial resolution, across multiple frequency bands, in a routine and repeatable manner. Such a capability would provide vertically resolved observations of planetary atmospheres from the ground on timescales inaccessible to individual spacecraft missions, enabling direct comparisons between spacecraft measurements and long-term monitoring at the same physical quantities and spatial scales.

The following sub-sections describe the specific scientific questions in the atmospheres of Venus and Mars that can be addressed through such high-resolution limb observations with ALMA-2040.
Instrumental requirements such as angular resolution, frequency coverage, and dynamic range are summarized in Table~\ref{tab:summary} at the end of this document.\\

\subsection{Characterizing Atmospheric Circulation in the Middle Atmosphere}
A major advantage of limb sounding is its sensitivity to horizontal winds. Because the line of sight traverses the atmosphere along a long tangential path, horizontal winds can be directly measured through their Doppler shifts. Direct wind measurements have already been demonstrated with the current ALMA for Titan, Jupiter, and Saturn \citep{Lellouch2019, Cordiner2020, Benmahi2021, Benmahi2022}. For Venus and Mars, despite the substantial amount of observational data obtained by spacecraft, direct measurements of atmospheric winds have remained limited, with winds mostly inferred indirectly from cloud tracking or temperature gradients using thermal-wind or cyclostrophic balance approximations \citep[e.g.,][]{Piccialli2008, Horinouchi2017}. Submillimeter heterodyne spectroscopy therefore provides a unique means of directly measuring atmospheric winds on these terrestrial planets. Before the advent of ALMA, CO Doppler-shift observations of the atmospheres of Venus and Mars were conducted using both single-dish telescopes \citep[e.g.,][]{Lellouch1991,Lellouch1994,Rengel2008,Cavalie2008,Clancy2008,Clancy2012,Clancy2015} and millimeter interferometers, including the Plateau de Bure Interferometer (PdBI) and the Nobeyama Millimeter Array (NMA) \citep{Moreno2009, Moullet2012}. These previous studies provided spatially resolved measurements of line-of-sight wind velocities over specific altitude ranges (around 100~km for Venus and 60~km for Mars), but with a relatively coarse vertical resolution of approximately 10$-$15~km. ALMA2040 will extend this capability to vertically resolved, three-dimensional wind measurements.

\begin{itemize}
    \item \textbf{Venus:} A major outstanding question is the atmospheric circulation between approximately 70 and 130 km altitude. While the lower atmosphere is dominated by strong retrograde zonal winds (a.k.a.\,super-rotation), the circulation above the cloud-top remains poorly understood. A sub-solar-to-anti-solar (SSAS) circulation driven by the day–night thermal contrast is expected to contribute, but observations \citep{Moullet2012,Clancy2012,Clancy2015} and recent general circulation model (GCM) studies \citep[e.g.,][]{Martinez2026} suggest a more complex circulation involving atmospheric waves, including gravity waves propagating upward from the cloud layer and transporting angular momentum. High-resolution measurements of the vertically-resolved wind fields, combined with GCMs, would provide a direct test of these mechanisms and new constraints on the origin and maintenance of Venusian super-rotation, and its role in mediating the seasonal climate. 
    
    \item \textbf{Mars:} The atmospheric circulation of Mars under relatively quiescent conditions is reasonably well characterized, but its response to global dust storms remains poorly understood. During these events, atmospheric temperatures can change by several tens of kelvin and the global circulation can be substantially reorganized. The high optical depth of dust prevents visible and infrared observations from probing the atmosphere within major storms, whereas submillimeter and millimeter observations are much less affected by dust. The 2018 global dust storm provided a demonstration of this capability through ALMA mapping of the wind field \citep{Kakinuma2025}. ALMA2040 will provide vertical resolution, allowing us to determine how dust-induced heating modifies the circulation at different altitudes.
\end{itemize}

CO will be the primary tracer for these Doppler-wind measurements on Venus and Mars. Its high volatility, long photochemical lifetime and abundance over a broad altitude range make it a useful tracer of atmospheric dynamics, while different rotational transitions, with different optical depths, probe different altitude ranges. Isotopologues such as $^{13}$CO and C$^{18}$O provide additional constraints. In particular, combining $^{12}$CO and $^{13}$CO can help disentangle temperature from CO abundance. Because adjacent CO rotational transitions are separated by approximately 115~\si{\giga\hertz}, the current ALMA generally observes only one transition per Band, limiting simultaneous coverage of a broad altitude range. ALMA2040, with coordinated observations across multiple Bands, could combine multiple transitions and isotopologues to retrieve vertically resolved wind and temperature structures over a broad altitude range.

\begin{figure}[t]
\centering
\includegraphics[width=0.9\textwidth]{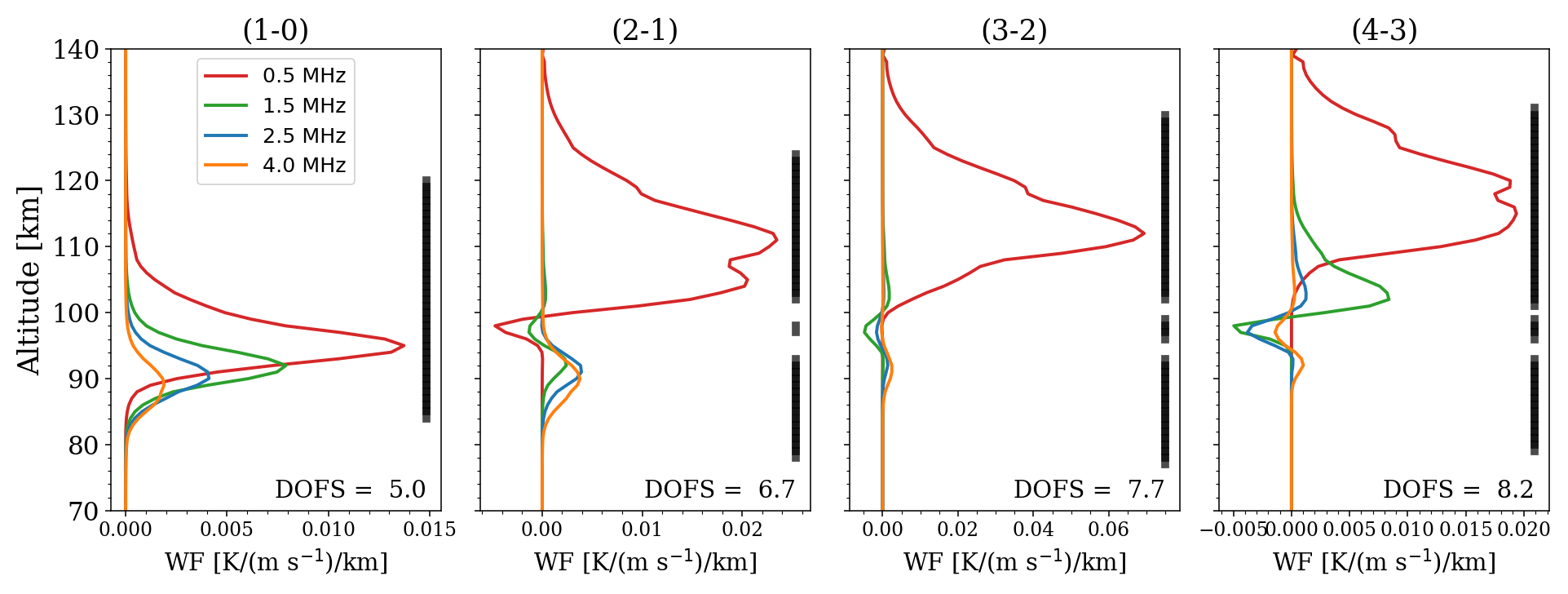}
\caption{\textbf{Sensitivity to line-of-sight winds in limb-viewing observations of Venus.} See the main text for details.}\label{fig:COsensitivity}
\end{figure}

Figure~\ref{fig:COsensitivity} shows the sensitivity of limb-viewing CO observations to line-of-sight winds. We simulated observations of the 
$^{12}$CO 1$-$0, 2$-$1, 3$-$2, and 4$-$3 lines at a tangent altitude of 60~km, assuming a spatial resolution of 0.01\arcsec, and a measurement noise of 1~K (for $\sim$1~\si{\mega\hertz} width). 
Each panel shows the weighting functions, which describe how the observed spectrum responds to small changes in wind speed at different altitudes. The functions are evaluated at frequency offsets of 0.5, 1.5, 2.5, and 4.0~\si{\mega\hertz} from the line center. The black bars along the right edge of each panel indicate the altitude ranges over which wind speeds can be retrieved from the full spectrum. The number of vertically independent pieces of information expected from the observed spectrum is quantified by the degrees of freedom for signal (DOFS). 
For example, with the 1$-$0 line, wind speeds can be retrieved over $\sim$85$-$120~km with five independent vertical layers. Higher rotational transitions provide stronger line signals and improve sensitivity at higher altitudes. The stronger signal near the line center also enhances the effective vertical resolution. 


\subsection{Comprehensive Observations of Trace Species and Isotopologues}
As already mentioned, a major advantage of limb sounding is its high sensitivity to trace atmospheric species enabled by the long atmospheric path length. ALMA2040 will exploit this sensitivity together with high vertical resolution to systematically observe trace molecules and isotopologues that have so far been poorly characterized, providing new constraints on the atmospheric chemistry and water cycles of Venus and Mars. Three particularly important science cases are described below. 

\begin{itemize}
    \item \textbf{Chemical Stability of CO$_2$-Dominated Atmospheres:} CO$_2$, the dominant atmospheric constituent of Venus and Mars, is photodissociated into CO and O by solar ultraviolet radiation. However, the direct recombination of CO and O to form CO$_2$ is spin-forbidden and therefore inefficient, implying that additional pathways are required to maintain a stable CO$_2$ atmosphere over long timescales. Catalytic cycles involving ClO$_x$ on Venus and HO$_x$ on Mars have been proposed as possible pathways, but their detailed reaction mechanisms remain poorly constrained \citep[e.g.,][]{McErloy1972,Demore1982,Luo2026}. \\
    A key to resolving this long-standing problem is the systematic observation of molecular species associated with these catalytic cycles, including HCl, ClO, H$_2$O, HO$_2$, and H$_2$O$_2$. Their abundance profiles can be compared with photochemical models to place observational constraints on the reaction pathways governing the production and loss of CO$_2$. Because photochemical processes depend strongly on local time, repeated observations over a range of local times will be particularly important. Venus offers a unique opportunity in this regard, as its changing geometry allows observations of different local-time regions from the dayside near superior conjunction to the nightside near inferior conjunction. 

    \item \textbf{The Fate of Water and HDO/H$_2$O Isotopic Fractionation:} Understanding how water was lost from the surfaces of Venus and Mars is central to reconstructing the long-term evolution and past habitability of terrestrial planetary environments. The vertical distribution and temporal variability of the HDO/H$_2$O ratio provide important diagnostics of this process. Infrared observations from spacecraft have revealed significant HDO/H$_2$O fractionation in the upper atmospheres of both planets \citep[e.g.,][]{Villanueva2022,Mahieux2024}, but independent measurements at other wavelengths and observations of its temporal variability are needed. \\
    This is particularly important for Mars, where heating of the upper atmosphere and enhanced upward transport of water vapor have been suggested during global dust storms. Monitoring the vertical distribution of HDO/H$_2$O before, during, and after such events would provide new constraints on the atmospheric processes governing the loss of water from Mars. Although H$_2$O itself is difficult to observe from the ground, the relatively high HDO/H$_2$O ratios on Venus and Mars make HDO a favorable tracer. In particular, the 893~\si{\giga\hertz} HDO line offers the prospect of high-precision ground-based observations. 

    \item \textbf{Venusian Sulfur Chemistry and Cloud Formation:} The formation and maintenance of the sulfuric-acid clouds that characterize Venus also remain poorly understood. SO$_2$, a key precursor of the clouds, exhibits substantial temporal variability whose origin is not yet understood. Moreover, its abundance decreases by 4$-$5 orders of magnitude between the lower atmosphere and the cloud-top region, a behavior that is not fully reproduced by current photochemical models and points to an unidentified SO$_2$ loss process \citep{Bierson2020,Lefevre2026}.\\
    ALMA2040 can constrain this missing sulfur chemistry by precisely measuring the vertical distributions of SO and SO$_2$ from the lower to the upper cloud layers. Sulfur chemistry may also be closely connected to the origin of the unidentified UV absorber in the Venusian clouds, which has been suggested to involve sulfur-bearing compounds. Monitoring the spatial and temporal variability of SO and SO$_2$ would therefore provide constraints not only on the sulfur cycle, but also on the formation and maintenance of the clouds and the nature of the unidentified UV absorber. \\
    SO$_2$ in the lower atmosphere has been observed in the near-infrared \citep{Marcq2021}; SO$_2$ in the upper cloud layer has been monitored over long time periods using high-resolution infrared spectroscopy \citep[][and subsequent references in the series by Encrenaz~et~al.]{Encrenaz2012}; and SO$_2$ and SO in the upper atmosphere, above the clouds, are frequently observed at submillimeter wavelengths \citep{Sandor2010,Piccialli2017}. The main observational gap lies in the altitude range from within the cloud layer down to the cloud base. For SO$_2$ in this altitude range, the only observational data available to date come from in-situ measurements obtained through past probe missions \citep{Bertaux1996}.\\
    Observations in the millimeter regime, particularly ALMA Band 2, are well suited to probing atmospheric regions from within the clouds to below the cloud layer because the clouds become increasingly transparent at these wavelengths. However, SO$_2$ absorption lines in this frequency range are very weak and strongly pressure-broadened, producing very broad and weak absorption features extending over several \si{\giga\hertz} (Fig.~\ref{fig:SO2band2}). Detecting these features therefore requires accurate and stable spectroscopy over a wide bandwidth. In addition to high sensitivity and angular resolution, a broad instantaneous bandwidth and high spectral accuracy, particularly stable bandpass calibration, will thus be important technical requirements for investigating sulfur chemistry on Venus with ALMA2040. It should be noted that, for this scientific objective, nadir observations are preferable to limb observations because they are less affected by CO$_2$ collision induced absorption opacity.
    \begin{figure}[h]
    \centering
    \includegraphics[width=0.48\textwidth]{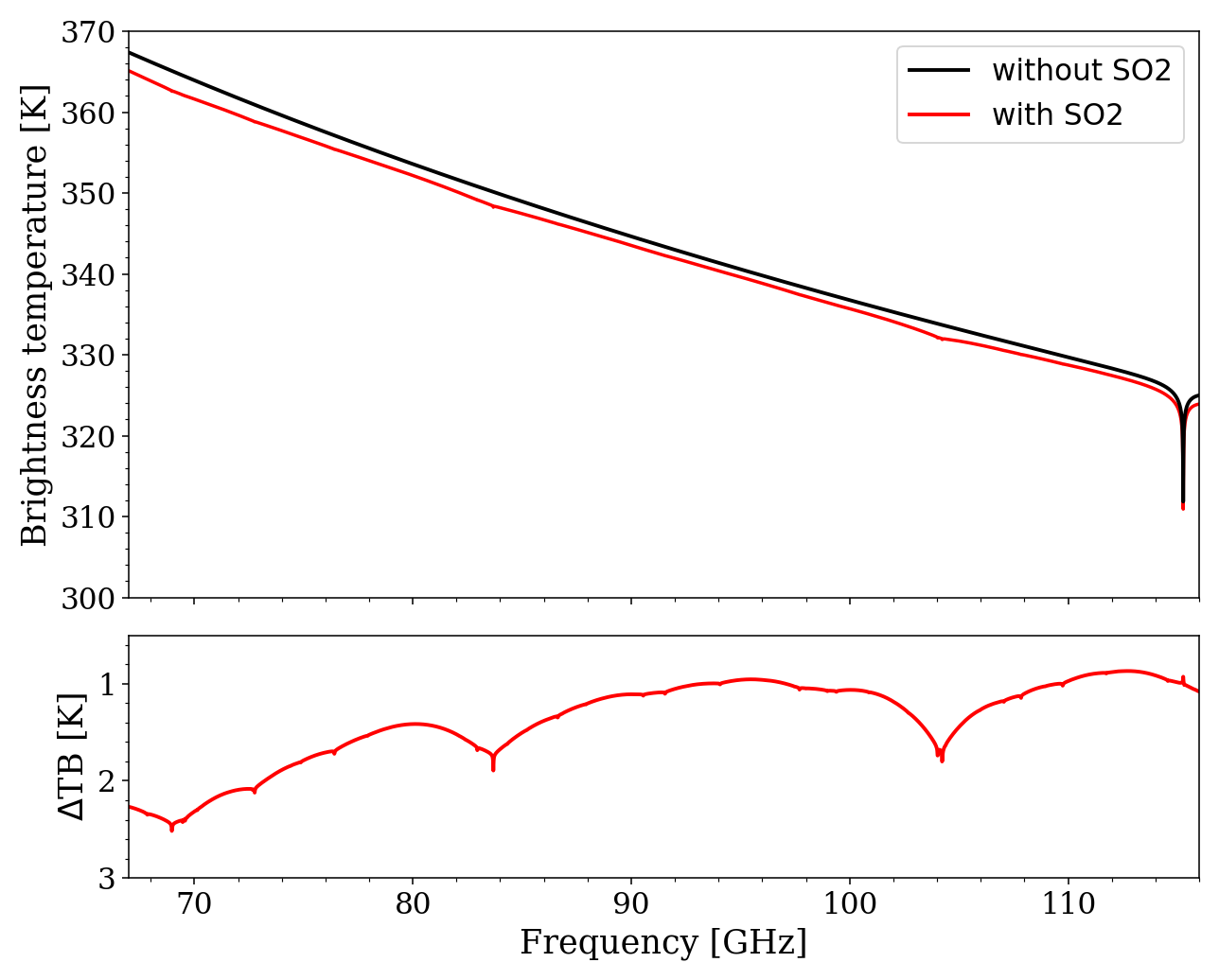}
    \caption{\textbf{Simulated nadir spectra of Venus at Band 2.} Spectral resolution is set to 16~\si{\mega\hertz}. The sharp absorption feature at 115~\si{\giga\hertz} is CO. Brightness-temperature difference between without- and with-SO$_2$ is shown in the lower panel.}\label{fig:SO2band2}
    \end{figure}
\end{itemize}

\section{Technical Considerations for the Maximum Recoverable Scale}
A major challenge in observing planets in the solar system with an interferometer is its limited sensitivity to extended emission. This limitation arises because short-baseline UV data cannot be obtained sufficiently, resulting in the so-called ``spatial filtering effect'' or ``resolve-out.'' 
For example, in the current ALMA configuration, each antenna array is designed to recover emission structures up to approximately 10 times the spatial resolution, a quantity referred to as the maximum recoverable scale (MRS). 

Accurately determining the vertical structure of the atmospheres of Venus and Mars through limb observations requires observations that cover the entire vertical extent of the atmosphere, which is approximately 100~km from the surface to the top of the atmosphere. Given the 5~km spatial resolution requirement for ALMA2040 described in this paper, this corresponds to a dynamic range of approximately 20 in spatial scale, exceeding the factor of 10 provided by the current ALMA configuration. 

In conventional astronomical observations, this limitation can be mitigated by combining multiple antenna arrays with different baseline ranges, such as compact and extended arrays. This approach enables observations to sample UV baselines over a broad range, from short to long baselines. However, such a combination is difficult to apply to planetary atmospheres, which can change significantly over short timescales. In particular, even the apparent diameter of the target can change within a few months, making it difficult to combine data obtained at different times with different array configurations.

Therefore, ALMA2040 requires not only very long baselines to achieve high spatial resolution, but also sufficient coverage at shorter baselines to recover extended atmospheric emission. One possible approach is to enable simultaneous observations with a second antenna array, such as an expanded version of the current ACA. Another possibility is to use a large-aperture single dish as a total-power antenna to recover the short-spacing information missing from the interferometric observations \citep{Cordiner2024}.\\

\section*{Acknowledgements}
This work was developed as part of the broader ALMA2040 community effort. We acknowledge the Lorentz Center and NOVA for supporting the community workshops and coordination activities that helped initiate and shape the ALMA2040 White Paper process.\\



\begin{table}
\centering
\caption{\textbf{Required observation capability of ALMA2040.}}
\label{tab:summary}

\small
\setlength{\tabcolsep}{4pt}

\begin{tabularx}{\textwidth}{
    >{\raggedright\arraybackslash}p{0.24\textwidth}
    >{\centering\arraybackslash}p{0.23\textwidth}
    >{\centering\arraybackslash}p{0.08\textwidth}
    >{\centering\arraybackslash}p{0.36\textwidth}
}

\toprule
\textbf{Science case}
& \textbf{Target Species}
& \textbf{Bands}
& \textbf{Instrumental Requirements} \\
\midrule

\addlinespace[3pt]
\multicolumn{4}{l}{
\textbf{Vertically-resolved \& high-sensitivity limb sounding}} \\
\addlinespace[2pt]

\hspace{2mm}Line-of-sight winds
& $^{12}$CO, $^{13}$CO, C$^{18}$O, C$^{17}$O
& 6, 7, 8
& \multirow{3}{*}{\makecell[l]{
$\bullet$~0.01\arcsec~resolution with MRS $\geq$0.2\arcsec.\\
$\bullet$~$\sim$100~kHz resolution over $\sim$200~MHz\\ bandwidth for line center. \\
$\bullet$~$\sim$1~MHz resolution over $\sim$4~GHz band-\\
width to cover the broad line wings. }} \\

\addlinespace[2pt]

\hspace{2mm}CO$_2$ chemical stability
& \makecell[c]{Venus: HCl, ClO\\
Mars: H$_2$O, HO$_2$, H$_2$O$_2$}
& \makecell[c]{6, 7, 9\\
5, 8}
& \\

\addlinespace[2pt]

\hspace{2mm}Fate of water
& H$_2$O, HDO, H$_{2}$$^{18}$O
& 5, 7, 10
& \\

\addlinespace[18pt]
\multicolumn{4}{l}{\textbf{Global disk mapping with nadir-viewing geometry}} \\
\addlinespace[2pt]

\hspace{2mm}Venus sulfur inventory
& SO$_2$, SO
& 2, 3, 6, 7
& \makecell[l]{
$\bullet$~$\sim$1\arcsec\,resolution with MRS $\geq$10$-$20\arcsec.\\
$\bullet$~$\sim$4~MHz resolution over $\sim$10~GHz\\
bandwidths.\\
$\bullet$~High spectral dynamic range $\geq \sim$2000.\\
$\bullet$~High bandpass calibration accuracy.}
  \\

\bottomrule
\end{tabularx}
\end{table}

\vspace{.5cm}

\bibliographystyle{apj_mod}
{\bibliography{bibliography}}
\end{document}